\input phyzzx
 
\def\dg{\dagger} 

\Pubnum={SHS-96-4\cr UUITP-08/96\cr}
\date={25 March 1996}
\titlepage
\title{Path Integrals and Parastatistics}
\bigskip
\author {Alexios P. Polychronakos\footnote\dagger
{poly@calypso.teorfys.uu.se}}
\address{Centre for Advanced Study, Norwegian Academy of Science and Letters,
\break 0205 Oslo, Norway}
\andaddress{Theoretical Physics Dept., Uppsala University\break
S-751 08 Uppsala, Sweden\footnote*{Permanent address}}
\bigskip
\abstract{The propagator and corresponding path integral for a
system of identical particles obeying parastatistics are derived.
It is found that the statistical weights of topological sectors
of the path integral for parafermions and parabosons are simply
related through multiplication by the parity of the permutation 
of the final positions of the particles. Appropriate generalizations
of statistics are proposed obeying unitarity and factorizability 
(strong cluster decomposition). The realization of simple maximal
occupancy (Gentile) statistics is shown to require ghost states.
}

\vfill
\endpage

\def\NP{{\it Nucl. Phys.\ }}
\def\PL{{\it Phys. Lett.\ }}

\def\PR{{\it Phys. Rev. \ }}
\def\PRL{{\it Phys. Rev. Lett.\ }}
\def\CMP{{\it Comm. Math. Phys.\ }}
\def\JMP{{\it J. Math. Phys.\ }}

\def\NC{{\it Nuovo Cimento \ }}

\REF\LM{J.M.\ Leinaas and J.\ Myrheim, \NC {\bf 37B}, 1 (1977).}
\REF\GMS{G.A.~Goldin, R.~Menikoff and D.H.~Sharp, \JMP {\bf 21},
650 (1980); {\bf 22}, 1664 (1981); \PR {\bf D28}, 830 (1983).}
\REF\WIL{F.~Wilczek, \PRL {\bf 48}, 1114 (1982); {\bf 49}, 957 (1982).}
\REF\WU{Y.S.~Wu, \PRL {\bf 52}, 2103 (1984).}
\REF\HA{F.D.M.\ Haldane, \PRL {\bf 67} (1991) 937.}
\REF\JAN{J.~Myrheim, {\it Anyons} (Notes for the Course on Geometric
Phases, ICTP, Trieste 6-17 Sept.~1993).}
\REF\GGG{O.W.~Greenberg {\it et al.}, hep-ph/9306225.}
\REF\GR{H.S.~Green, \PR {\bf 90}, 270 (1953).}
\REF\ST{O.~Steinmann, \NC {\bf 44}, A755 (1966).}
\REF\LS{P.V.~Landshoff and H.P.~Stapp, {\it Ann.~of Phys.} {\bf 45},
72 (1967).}
\REF\OC{Y.~Ohnuki and S.~Kamefuchi, \PR {\bf 170}, 1279 (1968); 
{\it Ann.~of Phys.} {\bf 51}, 337 (1969).}
\REF\DHR{S.~Doplicher, R.~Haag and J.~Roberts, \CMP {\bf 23}, 199 (1971);
{\bf 35}, 49 (1974).}
\REF\MG{A.M.L.~Messiah and O.W.~Greenberg, \PR {\bf B136}, 248 (1964);
{\bf B138}, 1155 (1965).}
\REF\HST{J.B.~Hartle and J.R.~Taylor, \PR {\bf 178}, 2043 (1969);
R.H.~Stolt and J.R.~Taylor, \PR {\bf D1}, 2226 (1970); \NP {\bf B19},
1 (1970); J.B.~Hartle, R.H.~Stolt and J.C.~Taylor, \PR {\bf D2},
1759 (1970).}
\REF\LDW{M.G.G.~Laidlaw and C.M.~DeWitt, \PR {\bf D3}, 1375 (1971).}
\REF\HAM{M.~Hammermesh, {\it Group Theory}, Addison-Wesley Eds.~(1962).}
\REF\CHA{S.~Chaturvedi, U.~Hyderabad preprint, hepth/9509150.}
\REF\SU{P.~Suranyi, \PRL {\bf 65}, 2329 (1990).}
\REF\KF{C.~Kostka, {\it Crelles Journal} {\bf 93}, 98 (1882);
H.O.~Foulkes, {\it Permutations}, Eds.~Gauthier-Villars, Paris (1974).}
\REF\GOV{A.B.~Govorkov, {\it Theor.~Math.~Phys.} {\bf 54}, 234 (1983).}
\REF\GRB{O.W.~Greenberg, \PRL {\bf 64}, 705 (1990).}
\REF\GEN{G.~Gentile, \NC {\bf 17}, 493 (1940).}
\REF\POL{A.P.~Polychronakos, \PL {\bf B365}, 202 (1996).}
\REF\MO{J.~Myrheim and K.~Olaussen, \PL {\bf B299}, 267 (1993).}
\REF\DHO{J.~Desbois, C.~Heinemann and S.~Ouvry, \PR {\bf D51}, 942 (1995).}
\REF\DV{A.D.~de Veigy, \NP {\bf B458} [FS], 533, 1996).}

Generalizations of the standard concepts of fermions and bosons 
have been extensively considered in the last few decades.
These can be broadly divided into ``phenomenological" (anyons [\LM-\WU],
exclusion statistics [\HA]), which are meant to give a better description
or understanding of systems of otherwise ordinary fermions or bosons, 
and ``fundamental," which would be genuinely new kinds of particles. 
A nice review of these approaches can be found in [\JAN], and a short 
and concise resum\'e of some relevant results can be found in [\GGG].

The first ever consistent extension of fundamental statistics, given
by Green [\GR], is parastatistics. In that, the standard bosonic 
or fermionic fields
which would create identical particles are replaced by composite
fields whose components commute with themselves and anticommute with
each other for parabosons, or vice versa for parafermions. The number
of components of the fields $p$ defines the ``order" of parastatistics.
In general, one can put at most $p$ parafermions in a totally symmetric
wavefunction, and at most $p$ parabosons in a totally antisymmetric one.
The degeneracies of occupation of more general multiparticle states
are in principle calculable but rather complicated. Parastatistics
in this approach has been well-studied [\ST-\DHR]. 

The above is a field theoretic realization of parastatistics. Just as
in the case of fermions or bosons, one can deal with a parastatistical
system at a fixed particle number in a first-quantized formalism. In
this approach, due mainly to Messiah and Greenberg [\MG,\HST], 
the $N$-body Hilbert space is decomposed into irreducible
representations (irreps) of
the particle permutation group $S_N$. Since the particles are 
indistinguishable, this group should be viewed as a ``gauge" symmetry of
the system, and states transforming in the same representation have to
be identified. Moreover, since all physical operators are required to
commute with the permutation group, each irreducible component is a
superselection sector. Therefore, one can project the Hilbert space
to only {\it some} of the irreps of $S_N$. Further, only one state in each
irrep need be kept as a representative of the multiplet of physically
equivalent states. The resulting reduced space constitutes a consistent
quantization of $N$ indistinguishable particles. The choice of included
irreps constitutes a choice of quantum statistics. In particular,
parabosons correspond to including only irreps with up to $p$ rows
in their Young tableau, while parafermions to ones with up to $p$ columns.
Clearly the cases $p=1$ reduce to ordinary fermions and bosons.

This description relies on a canonical quantization of the many-body system.
It is of interest to also have a path-integral formulation of a quantum
system, since this complements and completes the conceptual framework 
and usually offers orthogonal intuition in several cases. For ordinary
statistics this question was studied by Laidlaw and DeWitt [\LDW]. In this
paper, we provide such a realization for parastatistics, or, in general,
for any statistics where the Hilbert space is embedded in the tensor
product of $N$ one-particle Hilbert spaces (note that this excludes
anyons and braid statistics).

The starting point will be the coordinate representation of the full
(unprojected) Hilbert space, spanned by the position eigenstates
$|x_1 , \dots x_N > \equiv | x >$ (where $x_i$ can be in a space of any
dimension). The collection of such states for a set of distinct $x_i$
transforms in the $N!$-dimensional defining representation of $S_N$
$$
P |x> \equiv |Px> = |x_{P^{-1}(1)}, \dots x_{P^{-1}(N)} >
\eqn\Pdef$$
where $P$ is a permutation (the appearance of $P^{-1}$ in the above is 
necessary so that products of permutations be represented in the right
order). If any of the coordinates $x_i$ coincide the above is not the 
full defining representation any more. The set of such states, however, is
of measure zero (the coordinate space assumed continuous) and thus they can
be safely ignored. (We assume that there are no interactions singular at 
coincidence points that might dynamically make such states of nonzero weight.)

Projecting the Hilbert space to an irrep $R$ of $S_N$ amounts to keeping
only linear combinations of states within this multiplet transforming in 
$R$, that is,
$$
|a;x> = \sum_P C_a (P) P |x> ~,~~~ a = 1, \dots d_R ~,~~d_R =dim(R).
\eqn\Ca$$
where the sum is over all elements of the permutation group and $C_a (P)$ 
are appropriately chosen coefficients. If we denote with $R_{ab} (P)$
the matrix elements of the permutation $P$ in the representations $R$,
$$
P |a,x> = \sum_b R_{ab} (P^{-1} ) |b,x>
\eqn\Pa$$
The defining representation decomposes into irreducible components,
classified by Young tableaux, each appearing with a certain multiplicity.
Should we keep only one irrep out of each multiplicity or the whole multiplet?
To decide it, note that if instead of the base state $|x>$ for the
construction of the states $|a,x>$ we choose a different permutation
$P_o |x>$, then although the new states $|a, P_o x>$ constructed through
\Ca\ still transform in the irrep $R$, in general they are {\it not}
linear combinations of $|a,x>$ but rather span a different copy of $R$.
Since we can continuously move in the configuration space from $|x>$ to
$P_o |x>$, we conclude that we must keep {\it all} irreps $R$ within each
multiplet. (In other words, although for each point in the Hilbert space
$|x>$ this multiplet is reducible, the fiber of these representations
over the Hilbert space is connected and irreducible.)

To realize explicitly the above, we construct the states
$$
|ab,x> = \sqrt{d_R \over N!} \sum_P R_{ab} (P) P |x>
\eqn\ab$$
Using the group property of the representation $R(P_1 ) R(P_2) = R(P_1 P_2 )$,
we deduce that under the action of the group $S_N$ and under change of
base point $x$ the above states transform as:
$$
P |ab,x> = \sum_c R_{ac} (P^{-1} ) |cb,x>~,~~~
|ab,Px> = \sum_c R_{cb} (P^{-1} ) |ac,x>
\eqn\LR$$
Thus we see that the first index in these states labels the different
elements of a single irrep $R$, while the second index labels the
different equivalent irreps in the multiplet. Since both indices take
$d_R$ values, we recover the standard result that each irrep of $S_N$
is embedded in the defining representation a number of times equal to
its dimension.

Consider now the matrix element $<ab,x| A |cd,y>$, where $A$ is any
physical operator, that is, any operator commuting with all elements $P$
of $S_N$. Substituting the definition \ab\ and using the unitarity
of $P$ ($P^\dg = P^{-1}$) and of $R$ ($R_{ab}^* (P) = R_{ba} (P^{-1} )$)
we obtain, after a change in summation variable,
$$
<ab,x| A |cd,y> = {d_R \over N!} \sum_{P,P',e} R_{be} (P' ) R_{ea} ( P^{-1} )
R_{cd} (P) <x|A P'|y>
\eqn\AA$$
Using further the orthogonality (Shur's) relation (see, e.g., [\HAM])
$$
\sum_P R_{ab} (P) R_{cd} (P^{-1} ) = {N! \over d_R} \delta_{ad} \delta_{bc}
\eqn\SO$$
we finally obtain
$$
<ab,x| A |cd,y> = \sum_P \delta_{ac} R_{bd} (P) <x|A|Py>
\eqn\AB$$
Let us first choose $A=1$. Then the above provides the overlap between
the states
$$
<ab,x|cd,y> = \sum_P \delta_{ac} R_{bd} (P) \delta (x - Py )
\eqn\Norm$$
For $x$ in the neighborhood of $y$ it is $P=1$ which contributes to
the normalization, for which $R_{bd} (1) = \delta_{bd}$ and we recover
the standard continuous normalization between the states. 

Now we can choose $A=e^{-i H t}$, where $H$ is the hamiltonian, and thus
find the propagator $G(ab,x;cd,y|t)$ between the states of the system. 
It is clear from \AB\ that the first index $a$ in the state $|ab,x>$
propagates trivially. Since this is the index that corresponds to the
different but physically equivalent states within each irrep $R$, we
conclude that the required projection of the Hilbert space to the physical 
subspace amounts to simply omitting this index from all states. (That is,
freeze this index to the same fixed value for all states of the theory;
no physical quantity will ever depend on the choice of this value.) On
the other hand, the second index, corresponding to different equivalent
irreps, does {\it not} propagate trivially and must, as argued before,
be kept. We are led therefore to the physical states $|ba,x> \to |a,x>$ 
and the propagator
$$
G_R (a,x;b,y|t) = \sum_P R_{ab} (P)\, G(x,Py;t)
\eqn\G$$
where $G(x,Py;t) =~ <x| e^{-iHt} P|y>$ is the usual many-body propagator.
We note that, due to the transformation property \LR, the states
$|a,Px>$ are linear combinations of states $|a,x>$. Therefore, projecting
down to the physical subspace corresponding to $R$ amounts to trading
the original $N!$ copies of physically equivalent states $|Px>$ for
a number $d_R$ of {\it global internal degrees of freedom} for the system,
labeled by the index $a$. 

It is now easy to write down the path integral corresponding to identical
particles quantized in the $R$-irrep of $S_N$. $G(x,Py;t)$ can be expressed
as an $N$-body path integral in the standard way, with particles starting 
from positions $x_i$ and ending in positions $Py_i = y_{P^{-1} (i)}$.
Since all permutations of particle
positions are physically equivalent, \G\ instructs us to sum over {\it all}
sectors where particles end up in such permuted positions, weighted with the
factors $R_{ab} (P)$ depending on the internal degrees of freedom of the 
initial and final states. From \ab, \Norm\ we can write the completeness 
relation within the physical subspace
$$
I_R = \int {d^N x \over N!} \sum_a |a,x> <a,x|
\eqn\Comp$$
and with the use of \Comp\ it is easy to prove that the above path integral
is unitary, that is, 
$$
\int {d^N y \over N!} \sum_b G(a,x;b,y|t) \, G(b,y;c,z|t') 
= G(a,x;c,z|t+t') 
\eqn\Unit$$

The extension to parabosons, parafermions or any similar statistics is
immediate. Let $S = \{ R_1 , \dots R_n \}$ be the set of allowed irreps
of $S_N$ in the Hilbert space. The internal degree of freedom now takes 
values $A=(R,a)$, where $R \in S$ and $a=1, \dots d_R$ labels the internal
degrees of freedom within each irrep. So, overall, $A$ takes 
$d_{R_1} + \cdots d_{R_n}$ different values.
The propagator (and corresponding path integral) is obviously
$$
G_S (A,x;B,y|t) = \sum_P S(P)_{AB} \, G(x,Py;t) ~,~~~
{\rm where}~~ S(P)_{AB} = \delta_{R_A , R_B } ({R_A})_{ab} (P)
\eqn\GS$$
For parabosons (parafermions) of order $p$, $S$ is the set of Young tableaux
with up to $p$ rows (columns). We note that the irreps for parafermions are 
the duals of those for parabosons (the dual of a tableau is the tableau with
rows and columns interchanged). In an appropriate basis, the representation
matrices of dual irreps $R, {\tilde R}$ are real and satisfy
$$
{\tilde R}_{ab} (P) = (-1)^P R_{ab} (P)
\eqn\dual$$
where $(-1)^P$ is the parity of the permutation. We arrive then at the 
relation between the weights for parabosons and parafermions of order $p$:
$$
S_{pF} (P)_{AB} = (-1)^P S_{pB} (P)_{AB} 
\eqn\FBq$$
This extends a similar relation for ordinary fermions and bosons, 
for which there are no internal degrees of freedom and $S_B (P) =1$.

From the path integral we can evaluate the partition function, by simply
shifting to the euclidean periodic propagator $G_E (\beta) = e^{-\beta H}$ and 
summing over all initial and final states, with the measure implied by 
\Comp. Given that
$$
\sum_a R_{aa} (P) = \tr R (P) = \chi_R (P)
\eqn\CH$$
we get the expression in terms of the characters of $S_N$
$$
Z_S (T) = \int {d^N x \over N!} \sum_P S(P) <x| G_E (\beta) |Px> ~,~~
{\rm where}~~ S(P) = \sum_{R \in S} \chi_R (P)
\eqn\ZS$$
The interpretation in terms of a periodic euclidean path integral
is obvious. The characters $\chi_R (P)$ are a set of integers, and thus
the ``statistical factors" $S(P)$ weighing each topological sector of the
path integral are (positive or negative) integers. In the case of
parabosons of any order $p$, however, we note that the statistical 
weights are {\it positive} (or zero) integers. The ones for parafermions
can be either positive or negative, as given by 
$$
S_{pF} (P) = (-1)^P S_{pB} (P) ~,~~~ S_{pB} (P) \ge 0
\eqn\FBq$$
We do not have a general formula for $S_{pB} (P)$ for arbitrary $p$.

From the above results we can derive the partition function for a gas
of parastatistical particles as well as the allowed occupancy of 
single-particle states. Consider a collection of non-interacting
particles, for which the hamiltonian is separable into a sum of
one-body hamiltonians $H = \sum_i H(x_i)$. Let the energy eigenvalues of 
the one-body problem be $\epsilon_i$ and the corresponding one-body
Boltzmann factors $z_i = e^{-\beta \epsilon_i}$.
Consider now a sector of the euclidean path integral characterized
by the permutation of final points $P$. It is clear that this path
integral $Z_P$ decomposes into a product of disconnected components,
characterized by the fact that the particle worldlines in each component
mix particles only within the same component. This means that, within each
component, particles mix under a cyclic permutation (since following the
worldline of each particle must successively lead to every other particle
in the component). Each element $P$ of $S_N$ is then decomposed into a
product of commuting cyclic permutations. The number of particles $n$ 
participating in each cyclic permutation constitute the {\it cycles} of 
$P$ (obviously $\sum n = N$). We are led, thus, to the fact that for 
noninteracting particles
$$
Z_P = \prod_{n \in cycles(P)} Z_n
\eqn\ZZ$$
The path integral $Z_n$ for a cyclic permutation of $n$ particles, on the
other hand, can be thought of as the path integral of a single particle
winding $n$ times around euclidean time $\beta$. This means that
$$
Z_n (\beta) = Z_1 (n\beta) = \sum_i z_i^n \equiv W_n [ z_i ] 
\eqn\Zn$$
and the corresponding expression for $Z_P$ becomes
$$
Z_P = \prod_{n \in cycles(P)} W_n [z_i ]
\eqn\Pn$$
The expression for the full partition function then becomes
$$
Z_S = \sum_{R \in S} \sum_P {1 \over N!}\, \chi_R (P) \prod_{n \in cycles(P)} 
W_n [z_i ]
\eqn\ZSn$$
We recognize the sum over $P$ in \ZSn\ as Frobenius' relation, connecting
the sum over the Schur functions $W_n [z_i ]$ to the characters of $SU(M)$
$\chi_R [ z_i ]$. We get the final result
$$
Z_S = \sum_{R \in S} \chi_R [z_i ] = 
\sum_{R \in S} {\det ( z_i^{N-1-j+\ell_j} ) \over \det ( z_i^{N-1-j} )}
\eqn\ZSCH$$
where $\ell_j$ is the length of the $j$-th row of the Young tableau
of $R$. This reproduces the result of Chaturvedi for the partition function
[\CHA] and an earlier result of Suranyi for $p=2$ [\SU]. 
We stress that the above result holds only for noninteracting particles.
For interacting particles the topologically disconnected components $Z_n$
of the path integral $Z_P$ are still dynamically connected and factorization
fails. One has to go back to the full expression \ZS\ for the partition
function in that case.

To find the degeneracy of states, we need to decompose $\chi_R [z_i ]$
appearing in \ZSCH\ in monomials $\prod_i z_i^{p_i}$
$$
\sum_{R \in S} \chi_R [z_i ] = \sum_{\{p_i\}} D[p_i ] \prod_i z_i^{p_i}
\eqn\RD$$
The coefficients $D[p_i ]$ of these monomials, called Kostka-Foulkes numbers
[\CHA, \KF], are non-negative integers which determine the degeneracy of the 
state with $p_i$ particles occupying each energy level $\epsilon_i$. To find
these integers in a systematic way, we use the following trick: consider that
the particles are bosons and have an internal degree of freedom transforming
in the fundamental of $SU(N)$ (in fact, $SU(M)$ with $M \ge N$ would also do).
Since under total permutation of particle coordinates and internal 
degrees of freedom the states must transform trivially,
we conclude that the irrep of the color $SU(N)$ for each state must be the
same as the irrep of the coordinate permutation group $S_N$ (meaning they 
have the same Young tableau and thus the same symmetries). A state
with $p$ particles in the same level $\epsilon_i$ transforms in the 
$p$-fold symmetric irrep of $SU(N)$. Therefore, a state with occupancies $p_i$
for each level transforms under the direct product of $p_i$-fold symmetric
irreps, one for each level $\epsilon_i$. Decomposing this product into 
irreducible
components, we will obtain each representation $R$ of $SU(N)$ a number of
times $D_R [p_i ]$. Each such irrep will transform under a similar irrep
of $S_N$ and thus will correspond to a unique physical state in the
quantization of $N$ identical particles in the representation $R$. Therefore,
the degeneracy $D[p_i ]$ can be found by summing the number of times
that each allowed irrep $R \in S$ appears in the direct product of 
symmetric irreps $p_i$, which can be found using standard $SU(N)$ Young 
tableaux composition rules. 
We also see that, if the internal degree of freedom group 
is chosen to be $SU(p)$ (where $p$ may be smaller that $N$), we will
only get irreps with up to $p$ rows. Therefore, we recover the known result
that parabosons of order $p$ can be viewed as bosons with an internal
$SU(p)$ symmetry, where we identify each irrep of $SU(p)$ as a unique
physical state [\OC,\DHR]. A similar construction can be repeated starting
from fermions instead of bosons. We recover a dual expression for the
degeneracies $D[p_i ]$, where now we form the direct product of $p_i$-fold
fully antisymmetric irreps of $SU(N)$, and a similar expression of 
parafermions of order $p$ as fermions with an $SU(p)$ internal symmetry.

As was argued in [\LS,\HST], parastatistics particles obey the cluster
decomposition principle, in the sense that the density matrix obtained
by tracing over a subset of particles which decouple from the system
can be constructed as a possible density matrix of the reduced system
of remaining particles. From \ZS, however, we see that the partition
function of two dynamically isolated sets of particles $N_1$ and $N_2$
does {\it not} factorize into the product of the two partition functions,
since the statistical weights $S(P)$ in general do not factorize into
$S(P_1 ) S(P_2 )$ when $P$ is the product of two commuting elements
$P_1$ and $P_2$. Equivalently, this means that the occupation degeneracy
$D[p_i ]$ does not factorize into the product of individual occupation
degeneracies for each level $\epsilon_i$. This has important physical
implications. If the two sets of particles are totally isolated, it
does not make sense to evaluate the partition function of the total
system, since the statistical distribution can never relax to the one
predicted by that partition function. The individual partition functions
of the subsystems are the relevant ones. If, however, the two sets are
only weakly coupled, then initially each set will distribute according
to its reduced partition function, but after some relaxation time (depending
on the strength of the coupling between the two sets) they will relax to
the joint distribution function, which, we stress, will not even
approximately equal the product of the individual ones. Thus, cluster
decomposition holds in an absolute sense but fails in a
more realistic sense. In contrast, fermions and bosons
respect cluster decomposition in both senses.

The obvious generalization of quantum statistics, based always on the
assumption that the many-body Hilbert state is embedded into the tensor
product of many one-body Hilbert spaces, is to generalize the set of
allowed irreps $S$ beyond the one relevant to parastatistics. We may,
however, further include more than one state for each included irrep of
$S_N$. This seems unmotivated, in view of the fact that such states
are physically indistinct, but it is certainly consistent. It could mean,
for instance, that the particles have some hidden internal degrees of
freedom accounting for the extra degeneracy, which are invisible to the
present hamiltonian but may become dynamically relevant later. The most
general situation, then, is that we include $C_R$ states from each irrep
$R$. The generalization of all previous formulae for this case is quite
immediate, $S(P)$ and $C_R$ being related by
$$
S(P) = \sum_R C_R ~\chi_R (P) ~,~~ C_R = {1\over N!} \sum_P S(P) \chi_R (P)
\eqn\CS$$
The case of distinguishable particles 
(``infinite statistics" [\DHR,\GOV,\GRB]),
in particular, is reproduced by accepting all states in each irrep, that
is, $C_R = d_R$. Since $R$ appears exactly $d_R$ times in the defining 
representation of $S_N$, $S(P)$ above becomes the trace of $P$ in that
representation. But all $P\neq 1$ are off-diagonal in the defining
representation, so we get $S_{inf} (P) = N! \, \delta_{P,1}$, recovering
the standard distinguishable particles result.

We summarize here by pointing out that the most general statistics of the
type examined here is parametrized by any of three possible sets of numbers.
The first is, as just stated, the number of states $C_R$ accepted for
each irrep $R$ of $S_N$. Since the irreps of $S_N$ are parametrized by
the partitions of $N$ (lengths of rows or the Young tableau), there are
as many $C_R$ as there are partitions of $N$. The second set is the
statistical weights $S(P)$ appearing in the partition function (euclidean
path integral). Clearly these weights are invariant under conjugation
of $P \to Q P Q^{-1}$, since this simply amounts to a relabeling of the
particle worldlines. Thus $S(P)$ depends only on the conjugacy class
of $P$, that is, the cycles of $P$. The possible sets of cycles are
the same as the partitions of $N$; so, again, the $S(P)$ are numbered
by partitions of $N$. Finally, we could use the degeneracy of a many-body
occupancy state $D[p_i ]$ as our definition. There are as many ways to
distribute particles in one-body states as there are partitions of $N$,
so this set also has the same number of elements as the previous two.

What are the restrictions or criteria to be imposed on the above parameters?
The first one is unitarity, that is, the existence of a well-defined Hilbert
space with positive metric. This requires that $C_R$ be non-negative (no 
negative norm states) integers (no ``fractional dimension" states). The other
will be what we call ``strong cluster decomposition principle," that the 
partition function of isolated systems factorize. This is a physical 
criterion, rather than a consistency requirement. To summarize:
$$
\eqalign{
&\bullet~{\rm Unitarity:}~ C_R ~{\rm non-negative~integers} \cr
&\bullet~{\rm Strong~cluster~decomposition:}~ S(P) = \prod_{n \in cycles(P)}
S(n) ~~{\rm or} ~~ D[p_i ] = \prod_i D( p_i ) }
$$ 
The strong cluster decomposition, in particular, implies the existence of a 
grand partition function, obtained (in the case of noninteracting particles) 
by exponentiating the sum of all {\it connected} path integrals ($P$ a 
cyclic permutation of degree $n$) with weights $S(n)/n$ ($1/n$ is the 
symmetry factor of this path integral, corresponding to cyclic relabelings
of the particles). The grand partition function will further factorize
into a product of partition functions for each level $\epsilon_i$. Thus,
$S(n)$ are cluster coefficients connected to $D(n)$ in the standard way 
$$
\sum_{p=0}^\infty D(p) z^p = 
\exp \left( \sum_{n=1}^\infty {S(n) \over n} z^n \right)
\eqn\DS$$
The above formula, in fact, provides the easiest way to relate 
$D[p_i ]$ and $S(P)$ in the general case (no strong cluster property):
simply expand the right-hand side of \DS\ in powers of $z$ and substitute
every term $S( n_1 ) \cdots S(n_k )$ with $S(n_1 , \dots n_k )$.
This gives $D(p)$. To find $D[p_i ] = D(p_1 , \dots p_k )$ simply evaluate
$D(p_1 ) \cdots D(p_k )$ using the above formula and again consolidate
each product $S( n_1 ) \cdots S(n_k )$ into a single $S(n_1 , \dots n_k )$.

If we assume that $S(1) = D(1) = 1$, then it is easy to verify that the
{\it only} solution of the above two criteria is ordinary fermions and
bosons. The situation is different, however, when $S(1) = D(1) = q >1$ (this
would mean, e.g., that the particles come a priori in $q$ different 
``flavors"). The possibilities are manifold. All these generalized statistics
share the following generic features:

$\bullet$ The degeneracy of the state where $n$ particles occupy different
levels is $q^n$. (Indeed, $D(1,1,\dots 1) = D(1)^n = q^n$.)

$\bullet$ If state $A$ can be obtained from state $B$ by `lumping'
together particles that previously occupied different levels, then
$D(A) \le D(B)$. (E.g., $D(3) \le D(2,1) \le D(1,1,1)$.)

\noindent
This second property is actually related to the (weak) cluster decomposition
as formulated in [\HST], which is obviously covered by the strong cluster
property.

The above possibilities include the obvious special cases of $q_1$
bosonic flavors and $q_2$ fermionic ones ($q_1 + q_2 = q$), for which
$S(n)=q_1 - (-1)^n q_2$, along with many other. As an example, we give
the first few degeneracies for many-particle level occupation for all 
statistics with $q=2$:

$D(1)=2, D(2)=4, D(3)=8$

$D(1)=2, D(2)=3, D(3)=6,5,4(B+B)$

$D(1)=2, D(2)=2, D(3)=4,3,2(B+F),1,0$

$D(1)=2, D(2)=1, D(3)=2,1,0(F+F)$

$D(1)=2, D(2)=0, D(3)=0$

The specific choices denoted by $B+B$, $B+F$ and $F+F$ are the ones
corresponding to two bosonic, one bosonic and one fermionic, and two
fermionic flavors respectively. The topmost statistics could be termed
``superbosons" and the bottom one ``superfermions" of order 2.
We also remark here that the ``$(p,q)$-statistics'' introduced in [\HST]
can be realized as particles with $p$ bosonic and $q$ fermionic
flavors, where we identify each multiplet transforming irreducibly under
the supergroup $SU(p,q)$ as a unique physical state.

Finally, we direct our attention to the first known attempt to generalize
the ordinary Fermi or Bose statistical mechanics, by Gentile [\GEN]. 
The rule is simply that up to $p$ particles can be put in
each single-particle level. This corresponds to $D(n)=1$ for $n \le p$,
and $D(n)=0$ otherwise. This has been criticized [\GGG] on the grounds that
fixing the allowed occupations for each single-particle state is not
a statement invariant under change of single-particle basis. It is clear
that, in the language of this paper, any statistics satisfying the
unitarity requirement is consistent and basis-independent. Therefore,
Gentile statistics must violate unitarity. Indeed, it is easy to check
that all weights $C_R$ for such statistics are integers (this is
generic for all statistics with integer $D(n)$), but not necessarily
positive. In the specific case of $p=2$, e.g., where up to double
occupancy of each level is allowed, the degeneracies of each irrep of 
$S_N$ (parametrized, as usual, by the length of Young tableau rows) 
up to $N=5$ are
$$
C_2 = C_{21} = C_{22} = C_{221} =1,~ C_{111} = C_{1111} = C_{2111} =-1,
~~{\rm else}~C_R =0
\eqn\Gent$$
We see that representations $111$, $1111$, $2111$ correspond to
{\it ghost} (negative norm) states and their effect is to subtract
(rather than add) degrees of freedom. We also remark that the path
integral realization of exclusion statistics exhibits both negative 
and fractional statistical weights, signaling breakdown of unitarity 
[\POL]. This is inconsequential in that case, since exclusion
statistics is valid only as a macroscopic (statistical) description
of some (interacting) systems of particles.

In conclusion, we have presented the many-body propagators and
corresponding path integrals of particles obeying parastatistics or
any other type of statistics based on irreps of the permutation group.
We argued that there are many possible unitary generalizations obeying
the strong cluster decomposition principle, although they all require
more than one flavor of particles. Several other directions of investigation
and open questions suggest themselves. To name a few, the statistical 
mechanics of such generalized statistics particles should be examined.
Also, it should be checked if they can be realized as particles 
with specific hidden internal symmetries and an appropriate projection
of the Hilbert space, in a fashion similar to parastatistics. Independently,
it would be interesting to see if Gentile statistics can be consistently
realized by introducing `benign' ghosts which account for the negative
norm states while decoupling from all physical processes, just as
in gauge theories. Finally, a similar analysis could be attempted for
generalized particle statistics in $2+1$ dimensions.
In fact, a similar treatment, based on the permutation group, has been
used to obtain perturbative results for anyonic particles [\MO-\DV].
It would be interesting to examine whether non-abelian irreps of the 
braid group, instead of the permutation group, could be considered.

\ack{I would like to thank J.~Myrheim for discussions.}

\refout
\end